\documentclass[a4paper,11pt]{article}
\usepackage{pos}
\usepackage{amsmath}

\def\ll#1#2{\tilde{\lambda}_{#1}.\tilde{\lambda}_{#2}}
\def\llss#1#2{\tilde{\lambda}_{#1}.\tilde{\lambda}_{#2}\,\boldsymbol{\sigma}_{#1}\boldsymbol{\sigma}_{#2}}
\def\vec#1{\boldsymbol{#1}}

\title{Multiquark bound states and resonances}

\author*[a]{Jean-Marc Richard}
\author[b]{Alfredo Valcarce}
\author[c]{Javier Vijande}

\affiliation[a]{IP2I-IN2P3-CNRS \& University Lyon 1\\
	4, rue Enrico Fermi, Villeurbanne, France}

\affiliation[b]{Departamento de F\'\i sica Fundamental, University of Salamanca,\\
E-37008 Salamanca, Spain}

\affiliation[c]{University of Valencia, IFIC (CSIC-UV)\\
	Valencia, Spain}

\emailAdd{j-m.richard@ip2i.in2p3.fr}
\emailAdd{valcarce@usal.es}
\emailAdd{javier.vijande@uv.es}

\abstract{We review the chromoelectric and chromomagnetic mechanisms that tentatively lead to stable or metastable multiquark configurations. An alternative interpretation of the dynamics is the quark interchange between hadrons, as illustrated in the case of the fully-charm systems. 
}

\FullConference{Contribution to the Proceedings of the HADRON 2025 Conference, Osaka University, Japan, March 27 - 31, 2025, ed.\ A.~Hosaka
}

 \tableofcontents

\begin{document}
	\maketitle
\section{Introduction}
There are many possible  flavor configurations for multiquark states:  $uu\bar u\bar u$ \dots $ud\bar s\bar c$, \dots $bb\bar b\bar b$, and, of course, even more for the pentaquark of hexaquark systems.  It is customary, when working on some model, to claim that other approaches predict too many multiquarks, following the rule that ``he who wants to drown his dog accuses him of rabies''\,\footnote{Qui veut noyer son chien l'accuse de la rage (Molière)}. Actually, the simple quark model gives very few bound  multiquark states and a limited number of resonances above their lowest  threshold for dissociation into two hadrons. Thus  special circumstances are required to produce interesting states. 

In Sec.~\ref{se:CE:CM}, we review the trends suggested by chromoelectric and chromomagnetic forces. 
In Sec.~\ref{se:quark-exc}, we shall discuss the somewhat complementary view of quark interchange, which results into a coupled-channel hadron-hadron picture.
The conclusions are presented in Sec.~\ref{se:outlook}.
\section{Bound states in the quark model}\label{se:CE:CM}
%
Let us start with the simplest Hamiltonian: two quarks and two antiquarks with the same mass $m$ and a pairwise interaction corresponding to a color-octet exchange
\begin{equation}\label{eq:H1}
	H=\sum_{i=1}^4 \frac{\vec p_i^2}{2\,m}-\frac{16}{3} \sum_{i<j} \ll{i}{j} \,v(r_{ij}) \, ,
\end{equation}
where $r_{ij}=|\vec r_j-\vec r_i|$, $v(r_{ij})$ is the quarkonium potential, and $\tilde\lambda_i$ the color operators acting on the $i^{\rm th}$ quark, with suitable change for an antiquark. There is no bound state~\cite{Richard:2018yrm}. Hence, a stable multiquark requires to implement some improvements 
in~\eqref{eq:H1}: unequal masses, a spin-dependent interaction, three-body forces, ...
%

In early studies of multiquark states, the attention was not on the spin-independent terms of~\eqref{eq:H1}, but rather on the chromomagnetic interaction (CM), see, e.g., \cite{Jaffe:2004ph}. The predictions were based on the diagonalization of operators of increasing refinement such as
\begin{equation}
	C_1=\sum_{i<j}\llss{i}{j}~,\quad
	C_2=\sum_{i<j}\frac{\llss{i}{j}}{m_i\,m_j}~,\quad
	C_3=\sum_{i<j} a_{ij}\,\llss{i}{j}~,
\end{equation}
where $m_i$ are the quark masses and $a_{ij}$ are  taken from ordinary hadrons. 
The operator $C_1$ corresponds, up to a factor $m^{2}$, to the SU(3)$_{\rm F}$ limit of $C_2$. It was used for the early speculations on the $H$ dibaryon~\cite{Jaffe:2004ph}. In both $C_1$ and $C_2$ missing is the expectation value of the spin-spin component of the potential, that is included (together with $1/m_i\,m_j$) in the coefficient $a_{ij}$ of the variant $C_3$ of the chromomagnetic model.  It is then assumed, as an educated guess, that $a_{ij}$ can be borrowed from ordinary mesons and baryons. For instance $a_{sq}$ is the same in $cc\bar s\bar q$ as in $csu$. 
But the short range correlations might differ from a baryon to another one, as pointed out by Cohen and Lipkin~\cite{Cohen:1981ut}, and be sizably smaller in multiquarks, as stressed by Oka, Shimizu and Yazaki~\cite{Oka:1983ku} and others. Yet the model $C_3$ is amazingly successful, see, e.g., H\o gaasen et al.~\cite{Hogaasen:2005jv}, Y.R.~Liu et al.~\cite{Liu:2019zoy},  etc.

%
Another way to modify~\eqref{eq:H1} as to produce bound states is to introduce different masses and rely on the chromoelectric interaction (CE). Already in the 80s, it was shown that a $QQ\bar q\bar q$ configuration governed by
\begin{equation}\label{eq:H2}
	H=\frac{\vec p_1^2}{2\,M}+\frac{\vec p_2^2}{2\,M}+\frac{\vec p_3^2}{2\,m}+\frac{\vec p_4^2}{2\,m} -\frac{3}{16}\,\ll{i}{j}\,v(r_{ij})~,
\end{equation}
has a bound state below the $Q\bar q+Q\bar q$ threshold if the mass ratio $M/m$ is large enough~\cite{Ader:1981db,Zouzou:1986qh,Heller:1986bt}. See, e.g., \cite{Richard:2018yrm} for a detailed discussion and further references. This scenario was confirmed by the discovery of the $T_{cc}^+$ almost at the threshold.  A clear binding is expected for a $bb\bar u\bar d$ four-quark state. 

The case of $QQ\bar q\bar q$ is rather unique, with one pair of heavy quarks interacting in the tetraquark and none in the threshold. Consider for instance
$QQ'\bar Q'' \bar q$ vs.\  $Q\bar Q''+Q'\bar q$: 
they have the same cumulated $G=(-3/16) \sum \ll{i}{j}=1$ among heavy quarks, but $G=1$ in the tetraquark is built more symmetrically than $G=1$ in the threshold, and this is not favorable to binding.

The configuration $QQQQqq$ has the same cumulated strength $G=(-3/16) \sum \ll{i}{j}$ in the heavy sector as $QQQ+Qqq$, but spread more symmetrically and thus such quadruply heavy dibaryon is penalized. In short, a purely chromoelectric and pairwise  interaction does not predict $QQQQqq$ to be bound. 

Concerning $cc\bar c\bar c$, a very naive reasoning would be that  more heavy constituents would lead to improved stability, but in the quark model with pairwise color-octet-exchange forces, the system is not stable. But resonances do seem to exist~\cite{Wang:2022yes}. 

More uncertain is the case of $QQQQ'Q'Q'$ fully-heavy dibaryons. They are not bound in constituent models~\cite{Richard:2020zxb}, but bound in some lattice calculations~\cite{Junnarkar:2019equ}. If the discrepancy is confirmed, it will indicate a limit of the modeling by simple potentials.


In the pure chromoelectric model~\eqref{eq:H2}, the critical mass ratio $M/m$ for binding is rather elusive. However, if one combines the CE $QQ$ interaction and the favorable CM  $\bar u \bar d$ interaction in spin 0 and isospin 0, then $QQ\bar u\bar d$ is predicted to be stable for reasonable values of the heavy quark mass. 

Hunting for new configurations in which CE and CM effects add coherently has been done by several groups. For instance Leandri and Silvestre-Brac~\cite{Leandri:1995zm} focused on cases where the CM interaction is optimally attractive, especially in configurations with light quarks, and address the question whether for some spin-color-flavor combinations this attraction, when combined to the CE part of the interaction, becomes strong enough to produce bound states lying below the lowest hadron-hadron threshold.

The discovery of the hidden-charm pentaquark \cite{ParticleDataGroup:2024cfk} has also motivated studies dealing with other pentaquark configurations.
In~\cite{Richard:2017una}, we pointed out that constituent models predict \emph{stable} $\bar c cqqq$ states with isospin $I=3/2$. The reason is that spin forces in $qqq$ with isospin 3/2 are repulsive for ordinary $\Delta$ baryons, but might well be attractive if $qqq$ is not in a color singlet. 


\section{Quark exchange}\label{se:quark-exc}
The mechanism of quark exchange or quark rearrangement has been often evoked for building multiquarks, for instance in the context of the large $N_c$ limit of QCD, where $N_c$ is the number of colors~\cite{Lucha:2021mwx}. However, except for nucleon-antinucleon annihilation into three mesons~\cite{Green:1984wna}, the rearrangement formalism has not been pushed up to a quantitative  estimate. Let us sketch what could be such approach in the case of the fully-charmed $cc\bar c\bar c$.

First we describe the ground state and first radial excitations of $c\bar c$ through the expansion
\begin{equation}\label{eq;ccbar1}
	\phi(r)=\sum_{i=1}^n \alpha_i\, \exp(-a_i\,r^2/2)~,
\end{equation}
with typically $n=3$.

For the four-body problem $(1,2,3,4)=cc\bar c\bar c$,  one introduces the Jacobi coordinates and color state made of two singlets
\begin{equation}
	\vec x=\vec r_3-\vec r_1~,\quad \vec y=\vec r_4-\vec r_2~,\quad
	\vec z=(\vec r_2+\vec r_4-\vec r_1-\vec r_3)/\sqrt2~,\quad
	|1\rangle=(13)_1\,(24)_1~.
\end{equation}
This channel is described by the spatial function
\begin{equation}\label{eq:quark-exc1}
\Psi(\vec x,\vec y,\vec z)=\phi(x)\,\phi(y)\,\sum_k \beta_k\,\exp(-b_k\,z^2/2)=\sum_{i,j,k} \gamma_{ijk}\,\exp[-(a_i\,x^2+ a_j\,y^2+b_k\,z^2)/2]~.
\end{equation}
%
in which the intra-meson range parameters $a_i$ are frozen and the inter-meson one $b_k$ are varied. But with this single channel, there is no constraint along the $\vec z$ coordinate. 

One thus introduces the companion non-orthogonal channel in color state $|1'\rangle$ and the corresponding wavefunction
\begin{equation}\label{eq:quark-exc3}
	|1'\rangle=(14)_1\,(23)_1~,\qquad
	\Psi\,|1\rangle + \Psi'\,|1'\rangle
\end{equation}
where $\Psi'=\Psi(\vec x',\vec y',\vec z')$
%
and $\{\vec x', \vec y', \vec z'\}=\{\vec x,\vec y, \vec z\}_{3\leftrightarrow4}$.
%

In~\eqref{eq:quark-exc3}, there is a non-trivial effective interaction along the $\vec z$ variable, generated by the component of color $|1'\rangle$, and vice-versa for the interaction along $\vec z'$. 

If the wavefunction~\eqref{eq:quark-exc3} is used in connection with real scaling, one obtains promising plateaus near 7\,GeV. It can be refined by adding to the expansion~\eqref{eq:quark-exc1}  terms in $\vec x.\vec y$ in $\psi$  and $\vec x'.\vec y'$ and $\Psi'$, to account for the P-wave excitations of the $c\bar c$ subsystems. We envisage to push the investigations as to include spin-dependent forces, $(c\bar c)(c\bar c)(c\bar c)$ already studied by another method in~\cite{Xu:2025oqn}), and unequal masses, namely $(Q\bar q)(Q\bar q)$. In the latter case, it is not really an alternative to the chromoelectric model evoked in Sec.~\ref{se:CE:CM}, but another interpretation: due to strong $QQ$ attraction, the two $Q\bar q$ mesons better overlap in~\eqref{eq:quark-exc3} and build a coherent superposition. The mechanism of quark exchange has been stressed recently in connection with the $J/\psi\,J/\psi$ resonances \cite{Huang:2024jin}.

\section{Outlook}\label{se:outlook}

The constituent quark model, despite its simplicity, continues to offer a useful framework to identify favorable conditions for multiquark binding. But it clearly does not predict an overwhelming number of stable exotic states.

Still, some channels stand out. The interplay between CE attraction among heavy quarks and CM coherence among light ones has proven particularly promising, especially for systems like $QQ\bar u\bar d$, where both mechanisms act constructively.

Future work should go beyond qualitative models and aim at quantitative, fully dynamical calculations. This includes solving the full few-body problem with spin-dependent interactions and proper antisymmetrization, and also exploring the alternative formulation of quark exchange mechanisms and coupled-channel dynamics to better understand near-threshold resonances. This latter approach reconciles to some extent the simple quark model and the molecular picture. 

Moreover, lattice QCD results (especially for fully-heavy dibaryons) seem to challenge the constituent picture, and  point toward missing elements or limitations of pairwise potential models.

Experimentally, the search for multiquarks should remain guided by configurations that combine favorable mass ratios, color-spin couplings, and decay signatures. Pentaquark states with hidden charm or beauty, and double-heavy tetraquarks, remain top candidates, though others may surprise us.

\paragraph{Acknowledgments:}
We would like to thank Atsushi Hosaka and his colleagues for the  efficient organization of the conference HADRON 2025 at the University of Osaka in a stimulating and friendly atmosphere. 

%
%



\providecommand{\href}[2]{#2}\begingroup\raggedright\endgroup
	
\end{document}